\documentclass[10pt,prl,aps,twocolumn,showpacs,superscriptaddress]{revtex4}
\usepackage{graphicx}
\usepackage{amssymb}
\usepackage{bm}

\newcommand{\bc}{\begin{center}}
\newcommand{\ec}{\end{center}}
\def\ba#1{\begin{array}{#1}\displaystyle}
\newcommand{\ea}{\end{array}}
\newcommand{\z}{\\ \displaystyle}
\newcommand{\beq}{\begin{equation}}
\newcommand{\eeq}{\end{equation}}
\newcommand{\beqa}{\begin{eqnarray}}
\newcommand{\eeqa}{\end{eqnarray}}
\newcommand{\no}{\nonumber}

\newcommand{\bi}{\begin{itemize}}
\newcommand{\ei}{\end{itemize}}
\def\mato#1{\left(\ba{#1}} 
\def\matf{\ea\right)}

\def\lt#1{\left#1}
\def\rt#1{\right#1}

\def\b#1{\bar{#1}}
\def\frc#1#2{\frac{#1}{#2}}

\newcommand{\p}{\partial}
\newcommand{\Pexp}{{\cal P}\exp}

\newcommand{\bra}{\langle}
\newcommand{\ket}{\rangle}

\newcommand{\Tr}{{\rm Tr}}

\newcommand{\Or}{{\cal O}}

\newcommand{\ep}{\epsilon}

\newcommand{\cur}{{\cal J}}

\begin{document}

\title{New method for studying steady states in quantum impurity problems: The interacting resonant level model}

\author{Benjamin Doyon}
\affiliation{Rudolf Peierls Centre for Theoretical Physics, Oxford University, OX1 3NP, United Kingdom}

\date{\today}


\begin{abstract}
We develop a new perturbative method for studying any steady states of quantum impurities, in or out of equilibrium. We show that steady-state averages are completely fixed by basic properties of the steady-state (Hershfield's) density matrix along with dynamical ``impurity conditions''. This gives the full perturbative expansion without Feynman diagrams (matrix products instead are used), and ``re-sums'' into an equilibrium average that may lend itself to numerical procedures. We calculate the universal current in the interacting resonant level model (IRLM) at finite bias $V$ to first order in Coulomb repulsion $U$ for all $V$ and temperatures. We find that the bias, like the temperature, cuts off low-energy processes. In the IRLM, this implies a power-law {\em decay} of the current at large $V$ (also recently observed by Boulat and Saleur at some finite value of $U$).
\end{abstract}

\maketitle

Impurity models describe mesoscopic quantum objects in contact with large conducting leads. Quantum dots are much-studied examples, and experiments in a bias voltage \cite{experiments} lead to accurate descriptions of their non-equilibrium steady-state properties. In impurity models, such states are of high interest as they pose the theoretical challenge of capturing the effect of non-equilibrium in a truly quantum system, yet they constitute the simplest non-equilibrium situation, where properties are time independent. At low energies, interactions between the leads' Landau quasi-particles and the impurity occur in the $s$-wave channel, and the spectrum may be linearised around the two Fermi points. Then, the universal behaviour is described by free massless fermions on the half line (bulk conformal field theory) with a non-conformal boundary interaction at the end-point.  Many methods are known for studying equilibrium behaviors in such models, but understanding and accessing properties of non-equilibrium steady states is still a much harder task. Wilson's picture does not apply, and, for instance, the effect of a bias on low-energy processes (which determines the large-bias current) is still under study \cite{RoschKW01,DoyonA06}.

Two calculational schemes exist: the real-time Schwinger-Keldysh formulation and the scattering-state Lippman-Schwinger formulation. The former relies on the infinite extent of the half line in order to absorb the energy released by relaxation from an appropriate ``uninteracting'' density matrix to the steady state. It requires relaxation mechanisms (see, e.g. \cite{DoyonA06}), whose absence may lead to pathologies in perturbative expansions \cite{ParcolletH02}. The latter describes directly the expected end result just from ``how the state looks'' asymptotically far from the impurity. Hershfield's $Y$ operator \cite{Hershfield93} (see also the studies \cite{SchillerH98, DoyonA06, Han06}) gives a ``steady-state density matrix'' that encodes these scattering states. This is interesting, since a non-equilibrium steady state is not described by the usual density matrix, but it is still hard to apply to interacting systems. Exact results for integable models occur when the exact quasi-particles do not mix the biased particle baths \cite{FendleyLS95}. A recent proposal \cite{MehtaA05} suggested, in the interacting resonant level model (IRLM), a freedom in choosing the quasi-particles to get around this restriction, but the construction of Bethe ansatz eigenstates still needs justifications (for another recent study of the IRLM, see \cite{BordaVZ06}).

In this paper, we develop a new method for studying steady states in quantum impurity models from basic properties of Hershfield's $Y$ operator and ``impurity conditions''. Impurity conditions are part of the equations of motion that relate local operators on both sides of the impurity. In integrable models, they fix the scattering matrix of Bethe ansatz, and they are used in \cite{BoulatS07} to describe the infrared fixed point of the IRLM. We use them to develop an efficient perturbation theory in the scattering-state formulation. They allow us to bring chiral fields (representing collective massless modes in the bulk) through the impurity to the side where Hershfield's steady-state density matrix looks like the usual equilibrium one. The steady-state condition is then solved iteratively in terms of equilibrium averages, giving the full perturbative expansion without Feynman diagrams, where the combinatorics is dealt with using matrices on the space of impurity operators.

This is an improvement over Keldysh perturbation theory -- there is no special time ordering; and over other approaches in the Lippman-Schwinger formulation -- there is no need for explicit scattering states or $Y$ operator. It gives full perturbative expansions with respect to marginal or marginally relevant operators, which is impossible by standard methods. The re-summed expansion is an equilibrium average that may lend itself to numerical techniques. The method is also related to ``equations of motion methods'' \cite{EOM}; physically motivated truncations give non-perturbative results. Conceptually, the method shows how basic properties of the steady-state density matrix are sufficient to uniquely determine quantum averages; this may help in understanding variational approaches out of equilibrium.

In the IRLM, we calculate the universal non-equilibrium current for all voltages and temperatures to first order in Coulomb repulsion $U\ge 0$ (we verify that Callan-Symanzik equations hold). We find that large differences between the leads' filling energies (determined by the voltage) and the impurity energy cut off low-energy processes, giving a power-law decay at large voltages. Such a decay was also observed recently at a particular non-zero value of $U$ \cite{BoulatSII}. We note that the Bethe ansatz eigenstates of \cite{MehtaA05} do not satisfy our impurity conditions, except at $U=0,\infty$ (where exact quasi-particles are free). Finally, we give an argument to obtain the current at $U=\infty$, and find agreement with \cite{MehtaA05} and with free-theory expectations.

{\bf The hamiltonian.} We use units where $e=\hbar=1$. The IRLM is described by a quantum field theory hamiltonian $H=H_0+H_I$. $H_0$ represents two baths of one-dimensional free massless, spinless fermions (with quantum fields $\psi_1(x)$ and $\psi_2(x)$) on the half line, and $H_I$ the interaction with the impurity, an energy level that can be singly-occupied, at the end of the half line. Since there is no back-scattering, we can ``unfold'' the two baths and get two chiral right-moving fermions on the line. Using $\sqrt{2} \psi_e = \psi_1+\psi_2,\,\sqrt{2}\psi_o =\psi_1-\psi_2$, the hamiltonian is
\beqa
    H_0 &=& -i\int dx\,\lt(\psi_e^\dag(x)\p_x \psi_e(x) +\psi_o^\dag(x)\p_x\psi_o(x)\rt) \label{HI} \\
    H_I &=& t(\psi_e^\dag(0) d + d^\dag \psi_e(0)) + U(n_e(0) + n_o(0))D + \ep_d D\no
\eeqa
where $D=d^\dag d$ and $n_{e,o}(x) = \psi_{e,o}^\dag(x)\psi_{e,o}(x)$. Operators satisfy canonical anti-commutation relations $\{\psi_e(x),\psi_e^\dag(x')\} = \{\psi_o(x),\psi_o^\dag(x') \}= \delta(x-x')$ and $\{d,d^\dag\}=1$, $d^2=(d^\dag)^2=0$. $U$ represents the Coulomb repulsion between the leads' fermions and an occupied impurity, $\ep_d$ is the impurity energy level, and $t$ is the hybridization, which essentially determines the number of current-carrying states available. It is useful to keep in mind that in the unfolded picture, $x<0$ means before interacting with the impurity, and $x>0$ means after.

{\bf Steady states.} In quantum impurity systems, steady states are also time-independent quantum states. They are defined by prescribing the baths to be, at $x<0$, in thermal equilibrium at temperature $T$ with a chemical potential $V$ associated to a local conserved charge $Q$ of $H_0$, and by asking for time-independence with respect to the dynamics given by the full hamiltonian $H$. That is, quantum averages $\bra \cdots \ket \equiv \Tr\lt(\rho \cdots\rt) / \Tr\lt(\rho\rt)$ are described by the density matrix \cite{Hershfield93} $\rho = \exp\lt[-(H - VY)/T\rt]$, where $Y$ looks like $Q$ before interacting with the impurity:
\beq \label{ssdef}\ba{l}
      \bra \Or_1(x_1)\Or_2(x_2)\cdots\ket \stackrel{x_1<0,x_2<0,\ldots}= \z
    \frc{\Tr\lt(\exp\lt[-\frc1T(H_0-VQ)\rt] \Or_1(x_1)\Or_2(x_2)\cdots\rt)}{\Tr\lt(\exp\lt[-\frc1T(H_0-VQ)\rt]\rt)}
    \ea
\eeq
($\Or_i(x)$ are local operators), and is conserved by the dynamics \footnote{In order to describe a non-equilibrium steady state, $Y$ is expected to be a non-local conserved charge.}: $[H,Y]=0$. As we will see, these two properties are sufficient to uniquely characterise steady-state averages. For our purposes, the charge $Q = (N_2 - N_1)/2$ with $N_i=\int dx \psi_i^\dag(x)\psi_i(x)$ and the current $\cur = -i[H,Q] = t {\rm Re}(id^\dag\psi_o(0))$ describe a bias voltage and the associated particle current.

{\bf The method.} {\em Impurity conditions.} Eigenstates of $H$ can be constructed by linear combinations of states of the form $\int dx dx'\cdots\; g(x,x',\ldots) \psi_{e,o}^\dag(x)\psi_{e,o}^\dag(x')\cdots\;(d^\dag)^{0,1}|0\ket$ with pseudo-vacuum $|0\ket$, $\psi_{e,o}(x)|0\ket =0,\;d|0\ket=0$ (a vacuum for both the fermionic Fock space ${\cal F}$ and the impurity space ${\cal I}$). The Hilbert space ${\cal H}$ comprises states up to an energy distance $\Lambda$ from the pseudo-vacuum -- this represents finite leads' bandwidths of order $\Lambda$, a part of our cutoff scheme. Because of the linear spectrum in the bulk, the first-quantised wave functions $g$ have finite jumps at the impurity site. Hence, $\bra v|\ \Psi(x)\ |w\ket$ has a jump at $x=0$, where $|v\ket,|w\ket \in {\cal H}$ and $\Psi(x)$ is an ultra-local field (no derivatives) at $x$ times impurity operators. Using $[H_0,\Psi(x)] = i\p_x \Psi(x)$ in $\bra v|\ [H,\Psi(x)]\ |w\ket$, the jump is, as an operator on ${\cal H}$ (with $a<0, b>0$):
\beq\label{impcond} \ba{l}
    \Psi(0^+) - \Psi(0^-) \stackrel{\bra v| \cdot |w\ket}= \z \qquad
    i\,\lt(\int_{a}^b dx - \int_a^{0^-}dx - \int_{0^+}^bdx\rt) [H_I,\Psi(x)]~.
\ea\eeq

Eq. (\ref{impcond}) has ambiguities: $H_I$ contains operators at $x=0$, where there are jumps. This is an artefact of the un-physical point-like nature of the impurity in the model, and is lifted by ``spreading'' it in a region around $x=0$ (another part of the cutoff scheme). Equivalently, one can ``resolve'' the hamiltonian, by only using limiting operators $\psi_{e,o}(0^+),  \psi_{e,o}(0^-)$. A natural symmetric resolving\footnote{Other symmetric resolvings give to the same model, with possibly different $t,U,\ep_d$.}, which we will use, is the replacements $\psi_{e,o}(0)\mapsto (\psi_{e,o}(0^+)+\psi_{e,o}(0^-))/2$ (and hermitian conjugate) in $H_I$. In a longer version of this work \cite{II} we will discuss these issues and construct the Bethe ansatz eigenstates of the resolved hamiltonian.

Taking, then, $\Psi(x) = \psi_e(x)$ in (\ref{impcond}) gives
\[
    \lt(1 + \frc{i U}2D\rt) \psi_e(0^+) -\lt(1-\frc{iU}2D\rt) \psi_e(0^-)\stackrel{|w\ket}= -itd
\]
and a similar equation holds with $\psi_e\mapsto\psi_o$ and $t\mapsto0$. From (\ref{impcond}), these hold only as operators on ${\cal H}$. However, it can be explicitly checked on Bethe ansatz eigenstates that they hold more generally as linear maps ${\cal H}\to{\cal F} \otimes {\cal I}$. This is the meaning of the equality symbol above. This fact may be due to integrability of $H$, and fixes the scattering matrix of Bethe ansatz eigenstates (different from that of \cite{MehtaA05} if $U\neq 0,\infty$). It also gives ``right-limits'' in terms of ``left-limits''. For instance, with $u= \frc{2iU}{2i-U}$,
\beq
    \psi_e(0^+) \stackrel{|w\ket}=-itd + \lt(1-iuD\rt)\psi_e(0^-) \label{impcond2}~.
\eeq
Other equations are obtained from (\ref{impcond2}) by pre-multiplication by $d,d^\dag$ and $\psi_e^\dag(0^+),\psi_e^\dag(0^-)$ and by hermitian conjugation and replacement $e\mapsto o, t\mapsto 0$. These are our impurity conditions; they stay valid when multiplied by local operators at $x\neq0$. What they tell us is how local operators that are just to the right of the impurity can be brought to its left only using the dynamics. They give for the current $\cur=t {\rm Re}(id^\dag \psi_o(0^-))$ on $\cal H$.

{\em Steady state conditions.}  Using impurity conditions, the steady state condition $\bra [H,\Psi(x)]\ket =0$ for $x<0$ can be written with all local operators to the left of the impurity. We write $\Psi(x) = d_j \Or(x)$ for some $j=1,2,3$ where we denote $b_1 =d,\,b_2=d^\dag,\,b_3=D$. Then, we have
\[
    -i\p_x \bra b_j \Or(x) \ket = iA_j\bra b_j\Or(x)\ket + \bra (c_{j} + \sum_i b_i E_{i,j}(0^-))\Or(x)\ket
\]
where $A_j$ is $\frc{t^2}2+i\ep_d$, $\frc{t^2}2-i\ep_d$, $t^2$ and $c_j$ is $-t\psi_e(0^-)$, $t\psi_e^\dag(0^-)$, $0$ for $j=1,2,3$, and
\beq\label{matE}
	E_{i,j}(x) = \mato{ccc}-un(x) &0& -t\psi_e^\dag(x) \\ 0&\b{u}n(x)&-t\psi_e(x) \\
	itu\psi_e(x)&it\b{u}\psi_e^\dag(x)&0 \matf_{i,j}
\eeq
with $n=n_e+n_o$. This is valid also with $\Or(x)\mapsto\Or(x+x_1)\cdots \Or(x+x_n)$ for $x+x_j<0$.

{\em Full perturbative expansion in $U$.} This differential equation can be integrated, choosing integration constants for a finite limit $x\to-\infty$ of $\bra b_j\Or(x)\ket$:
\beq\label{f}
    \bra b_j \Or(x) \ket = ie^{-A_jx} \int_{-\infty}^x dx' e^{A_jx'} \bra (c_{j}+\sum_i b_iE_{i,j}(0^-))\Or(x')\ket~.
\eeq
Solving (\ref{f}) by iteration, with $\sum_i b_i E_{i,j}(0^-) \Or(x')$ in place of $b_j \Or(x)$ at $x\to0^-$ and repeating, gives:
\beq \label{exp}\ba{l}
    \sum_{j=1}^3 \bra b_{j} \Or_j(0^-)\ket =  \sum_{n=0}^{\infty} i^{n+1} \int_{-\infty}^0 dx_0\cdots dx_n
    \bra c^T e^{Ax_n} E(x_n) \;\times\z \times\;
    \cdots e^{Ax_1} E(x_1+\ldots+x_n)e^{Ax_0}\b\Or(x_0+\ldots +x_n)\ket^0
    \ea
\eeq
where $A={\rm diag}(A_1,A_2,A_3)$, $c^T=(c_1,c_2,c_3)$, $E$ is the matrix (\ref{matE}), and $\b\Or=(\Or_1,\Or_2,\Or_3)^T$. All terms of this series are averages of local operators strictly to the left of the impurity. Hence (\ref{ssdef}) was used, and $\bra\cdots\ket^{0}$ means free equilibrium averages like on its right-hand side. This is our main result: we have obtained the full perturbative expansion. Essentially, the use of impurity conditions has simplified the combinatorics involved. For $U=0$, the first interation gives the current in closed form, in agreement with e.g. \cite{SchillerH98}.

The expansion (\ref{exp}) can be written using path-ordered integrals as $i \int_{-\infty}^0 dx\,\bra c^T \Pexp\int_{0}^x  dx'\,(-iE(x') +A) \,\b{\Or}(x) \ket^0$. This can be evaluated using appropriate {\em random variables} instead of operators, which may be useful for numerics. It can also be re-summed using Baker-Campbell-Hausdorff formula, $i\int_{-\infty}^0 dx \bra c^T e^{-i(H_0 +iA +E)x} \b\Or(0) e^{iH_0x}\ket^0$, where $E=E(0)$. This should allow a full account of the diagonal part of $E$ through bosonisation techniques and give a perturbative expansion in $t$.

{\em Divergencies.} The expansion (\ref{exp}) is plagued by linear and logarithmic divergencies as $\Lambda\to\infty$. Linear divergencies (from normal-ordering with respect to $H_0$ in $E_{i,j}(x)$, $n(x)\mapsto :n(x):$, and from collisions $:n(x):\, :n(x'):$ as $x\sim x'$ in (\ref{exp})) will be absorbed into a re-definition of $\ep_d$ \cite{II}. Logarithmic divergencies are accounted for by the renormalisation group -- a good alternative regularisation that gives them is simply to ``avoid'' the impurity by changing the upper integration limits in (\ref{exp}) to $-\Lambda^{-1}$.

{\bf Results.} We take $U>0$, and that regularisation. We find the current (divergent and finite parts as $\Lambda\to\infty$)
\[
    \bra \cur\ket = m \lt( I_{cs} + U \lt(\lt(I_{cc}+I_c \,m\frc{\p}{\p m}\rt)I_{cs} + B
			\rt) + O(U^2) \rt)
\]
where $m=t^2/2$, $I_{cc} = \frc14 \sum_{\pm,\pm'} F(0,m\pm i\ep_d \pm' iV/2) +L(\Lambda)$, $I_c =  \frc12 \sum_\pm F(0,2m \pm iV/2) + L(\Lambda)$, $I_{cs} = \frc1{4i} \sum_{\pm} F(m\pm i\ep_d +iV/2,m \pm i\ep_d - iV/2)$ and
\[\ba{l}
	B = mT^2\int_{-\infty}^0 dy \int_{y}^0 dz\,\frc{e^{m(y+z)}  \cos\ep_d(y-z)}{\sinh\pi T z} (y-z) \times \z \lt[
			\frc{\cos\frc{Vz}2\sin\frc{V(y-z)}2}{\sinh\pi T(y-z)} -
			 \frc{y \cos\frc{Vz}2 \sin\frc{Vy}2 + z\cos\frc{Vy}2\sin\frc{Vz}2}{(y-z)\sinh\pi T y} \rt] \no
	\ea
\]
with $L(\Lambda) = \frc1\pi\log\frc{2\Lambda}{\pi T}$ and, with $\psi(z) =d\ln\Gamma(z)/dz$, $F(a,b) = \frc1\pi \lt(\psi\lt(\frc{a}{2\pi T}+\frc12\rt) - \psi\lt(\frc{b}{2\pi T}+\frc12\rt)\rt)$.

The average current satisfies the Callan-Symanzik equation $\lt(\frc{U}\pi m\frc{\p}{\p m} - \Lambda\frc{\p}{\p\Lambda}\rt)\bra\cur\ket = 0$. This checks RG out of equilibrium to all orders in $m$. It means that the scaling limit where the bandwidth $\Lambda$ is greater than all other scales is taken with the parameter $T_K^{1+\frc{U}\pi} = m\Lambda^{\frc{U}\pi}$ fixed (to first order in $U$). This trades the microscopic coupling $m$ for the scale $T_K$ that characterises the scaling limit. The renormalised result is as above with the replacements $m\mapsto T_K^{1+\frc{U}\pi} \mu^{-\frc{U}\pi}$ and $\Lambda\mapsto \mu$, for some ``perturbative scale'' $\mu$. Since the parameter $U$ does not flow \cite{Schlottmann82,BordaVZ06}, the renormalised perturbative expansion is automatically valid for all $T,V,\ep_d,T_K\ll \Lambda$, i.e. in the universal regime. Note that the exponent $1+U/\pi+O(U^2)$ is in agreement with standard bosonisation results.
\begin{figure}
\mbox{\includegraphics[width=6.5cm,height=4cm,angle=0]{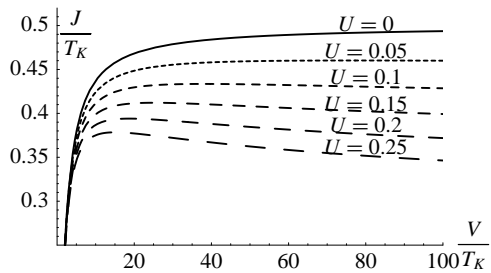}}\vspace{-0.4cm}
\caption{Universal current vs. voltage, $T_K$ fixed, $T=\ep_d=0$. A universal feature is the large-$V$ decay with a $U$-dependent power-law. At small $V$, its decrease with $U$ is not universal, and depends on the definition of $T_K$.}\vspace{-0.5cm}
\end{figure}

The full result is independent of $\mu$, which is only chosen to keep perturbative coefficients finite. Then, $\mu$ is the greater of $T_K$ or any physical infrared (IR) cutoff (quantities that do not allow arbitrarily many low-energy processes, like the temperature). It turns out that with $\mu\sim \sqrt{(\Delta_++T+T_K)(\Delta_-+T+T_K)}$, with $\Delta_\pm\equiv |V/2\pm\ep_d|$, when any of $T,T_K,\Delta_\pm$ is/are much larger, the first order perturbative coefficient stays finite.

Why are the energy separations $\Delta_\pm$ cutting off low-energy processes? Recall that transfers of particles occur around a point in space, so that by Heisenberg's uncertainty principle all energy scales in the leads may be involved. When at least one of $\Delta_\pm$ is high, there cannot be many low-energy transfers without many high-energy transfers on at least one side of the impurity. Here we find that each side contributes a factor of a square-root to the IR cutoff. That the voltage is a good infrared cutoff was also observed perturbatively in the Anderson \cite{Oguri02} and Kondo \cite{DoyonA06} models.

One implication is that the current vanishes as a power law as $V\gg T,T_K,|\ep_d|$ (see Fig. 1). At small temperatures, $V\gg T_K\gg T,|\ep_d|$, we have
$\bra\cur\ket \sim \frc12 T_K \lt(8 e^{\psi(1/2)}\frc{T_K}V\rt)^{\frc{U}\pi+O(U^2)} \lt(1 +  O(U^2)\rt)$. This is explained as follows. Observe that at $U=0$, the current saturates at large $V$, to a value proportional to the number of current-carrying hybridized states, roughly $m/\Lambda$. According to RG, the effect of $U\neq 0$ is that at large energies, $m/\Lambda$ decreases with a power law. Since $V$ is an infrared cutoff, at large $V$ the system is at large energies.  Note that if $|\ep_d|$ is kept of the same order as $V/2$, the power law changes to $V^{-\frc{U}{2\pi}}$.

On the other hand, the low-$T$ linear current, $T_K \gg T, V,|\ep_d|$, is (with $\b{T}\equiv \pi T/T_K$)
$2\pi \bra\cur\ket/V \sim 1+g_2\b{T}^2 + g_4\b{T}^4 + O(\b{T}^6)$ with $g_2= -\lt(16e^{2\psi(1/2)}\rt)^{-\frc{U}\pi}/3$ and universal ratio $g_4/g_2^2=21/5-U/\pi$, everything up to $O(U^2)$. The universal ratio is in agreement with IR conformal perturbation theory \cite{BoulatS07}.

Note that fixing $T_K$, the current {\em decreases} as $U$ is increased, for all voltages and temperatures (see Fig. 1 for the zero-temperature case). The fact that $T_K$ can be chosen in such a way is universal. The opposite is observed for finite but large enough bandwidth, finite voltage and fixed $t$, in agreement with \cite{BordaVZ06, MehtaA05}.

Physical arguments also give the current at $U=\infty$, where $u=2i$. Here, there cannot be a fermion at $x=0$ and an occupied impurity level simultaneously. Hence, both operators $d \,n(0^-),\, d^\dag n(0^-)$ must average to zero. In (\ref{f}) (with linear divergencies substracted), this brings the matrix $E$ to its $U=0$ form, so that the current is $m\,I_{cs}$, in agreement with \cite{MehtaA05}.

{\bf Conclusions.} We have developed an efficient method for studying steady states in impurity models, applied it to the current in the IRLM, and identified the non-trivial voltage-related quantities ($\Delta_\pm$) that cut off low-energy processes. It can be applied to any single-impurity model with linear spectrum and with left-/right-moving separation of the leads' quasi-particles (where unfolding can be done): Kondo and Anderson models, and impurity models where leads are Luttinger liquids (bosonic collective modes separate). In the non-equilibrium Kondo model with an impurity magnetic field, it may simplify the treatment of the pathologic perturbative expansion \cite{ParcolletH02}. Although our method may be restricted to integrable {\em dynamics} (this is to be clarified), it does not rely on integrability of the {\em steady state}; the non-equilibrium steady states in the Kondo and Anderson models may not be integrable. It would be very interesting to develop efficient, out-of-equilibrium numerical methods from the re-summed perturbative expansion obtained here.

This work was supported by EPSRC post-doctoral fellowship GR/S91086/01. I am grateful to F. Essler and H. Saleur for enlighting discussions and continuous interest in this work; to N. Andrei, J. Cardy, O. Castro Alvaredo and A. Ludwig for encouragements, discussions and comments on the manuscript; and to T. Imamura for correcting two formulas in an earlier version.

\null

\end{document}